\documentclass{jpsj-suppl}
\usepackage{txfonts} 

\title{$K^-K^-pp$ - an important gateway toward multi-kaonic nuclei}

\author{Shuji Maeda$^{1}$, Yoshinori Akaishi$^{2,3}$, and Toshimitsu Yamazaki$^{2,4}$}

\inst{$^{1}$Department of Agro-Environmental Science, Obihiro University of Agriculture and Veterinary Medicine, Hokkaido, Japan\\
$^{2}$RIKEN, Nishina Center, Wako, Saitama 351-0198, Japan\\
$^{3}$ Institute for Particle and Nuclear Studies, KEK, Tsukuba, Ibaraki 305-0801, Japan
$^{4}$Department of Physics, University of Tokyo, 7-3-1 Hongo, Bunkyo-ku, Tokyo 113-0033, Japan}

\email{yamazaki@nex.phys.u-tokyo.ac.jp}

\recdate{April 3, 2016, HYP2015 Proc., in press}

\abst{Based on our Faddeev-Yakubowsky calculations of four-body kaonic nuclear systems it was revealed that the structure of $K^-K^-pp$ is well approximated by two $\Lambda^*=K^-p$'s with strong mutual attraction. It is vitally important to study this nucleus, as it is an essential gateway toward multi-$\Lambda^*$ nuclei. Two experimental proposals are presented: i) production of $K^-K^-pp$ by $p+p$ reactions at $T_p = 7$ GeV, and ii) search for $K^-K^-pp$ from the invariant mass of $M(\Lambda\Lambda)$ around 2.6 GeV/$c^2$ in high-energy HI reactions.  
}

\kword{$K^-K^-pp$, multi-$\Lambda^*$ nuclei, structure and formation in HI reactions}

\begin{document}
\maketitle

\section{Introduction}

The finding of high-density $K^-pp$ at DISTO \cite{Yamazaki:10} and at J-PARC E27 \cite{Ichikawa:15} raises great interest in future studies of double kaonic bound state $K^-K^-pp$ both theoretically and experimentally. Here we propose to investigate $K^-K^-pp$ from i) the direct production with $p + p$ reactions at 7 GeV and from ii) indirect production in high-energy heavy ion collisions.

\begin{figure}[tbh*]
\begin{center}
\includegraphics[width=1.\textwidth]{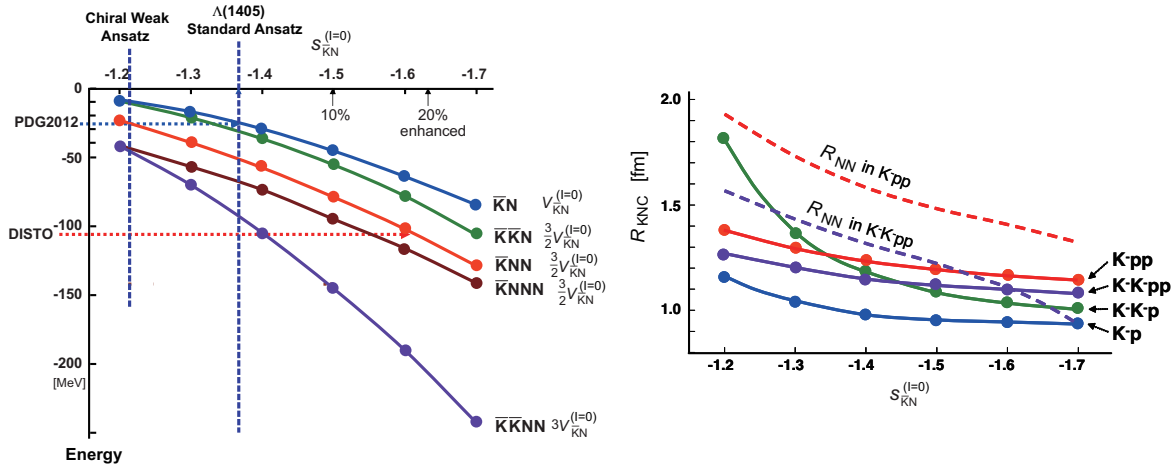}
\caption{Overview of the energy levels (Left) and the shrinkage (Right) of the calculated $K^-p$, $K^-pp$, $K^-K^-p$, $K^-ppn$ and $K^-K^-pp$ as functions of the $\bar{K}N$ interaction strength parameter $s_{\bar{K}N}^{(I=0)}$. From Maeda {\it et al.} \cite{Maeda:13}}
\label{fig:f1}
\end{center}
\end{figure}

\section{Comprehensive Calculations on $K^-p$, $K^-pp$, $K^-K^-p$, $K^-ppn$ and $K^-K^-pp$}

Extending our initial theoretical investigations and predictions \cite{Akaishi:02,Yamazaki:02} Maeda {\it et al.} \cite{Maeda:13} carried out Faddeev-Yakubovsky calculations for $K^-p$, $K^-K^-p$, $K^-pp$, $K^-ppn$ and $K^-K^-pp$ by varying the basic $(\bar{K}N)_{I=0}$ interaction in a wide range, thus achieving an overview of the binding energies and shrinkage, as shown in Fig.~\ref{fig:f1}.

The energy levels of typical few-body kaonic nuclei starting from $K^-p$, as we add nucleons and/or a $K^-$, are shown in Fig. 2. For each starting ansatz for $K^-p$, "$\Lambda^*(1420)$", "PDG" and "DISTO",  evolution of increasing binding energies with addition of a nucleon are seen. It is remarkable that the energy level of $K^-K^-pp$  drops down to -190 MeV for DISTO interaction that is compatible with the observed $K^-pp$ level at DISTO. On the other hand, the chiral predictions stay at shallow energy levels even for $K^-K^-pp$ \cite{Barnea:12}. 

\begin{figure}[tbh]
\begin{center}
\includegraphics[width=0.6\textwidth]{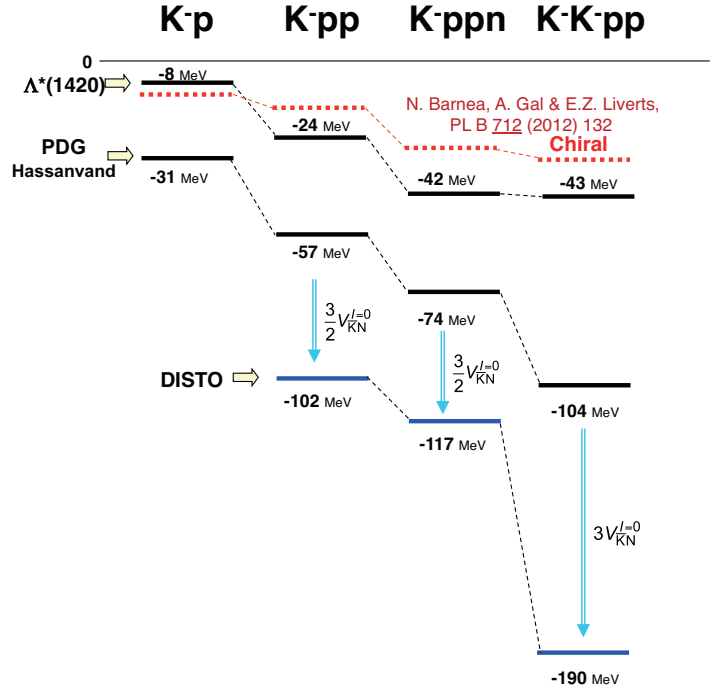}
\caption{Comparison of the predicted energy levels of $K^-p$, $K^-pp$, $K^-ppn$ and $K^-K^-pp$. The starting levels of $K^-p$ are indicated by $\Lambda^*(1420)$ including Barnea {\it et al.} \cite{Barnea:12}, PDG and DISTO that is compatible with the DISTO value of $K^-pp$ \cite{Yamazaki:10}. From Maeda {\it et al.} \cite{Maeda:13}. }
\label{fig:f12}
\end{center}
\end{figure}

\begin{figure}[tbh]
\begin{center}
\includegraphics[width=0.4\textwidth]{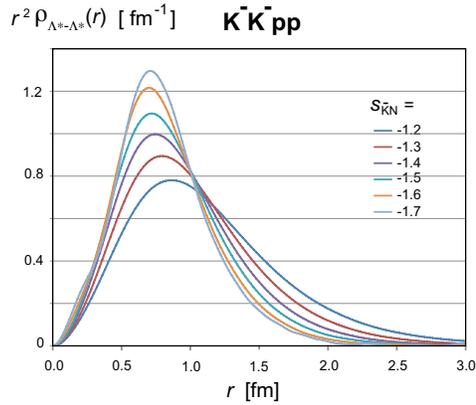}
\caption{(Left) Calculated distributions of the $\Lambda^*$-$\Lambda^*$ distance in $K^-K^-pp$ at various $\bar{K}N$ strength parameters. From \cite{Maeda:13}. (Right) An ensemble of $\Lambda^*$'s together with Heitler-London type covalent bonding of bosonic $K^-$'s that produces super-strong nuclear force \cite{Yamazaki:07a}. }
\label{fig:f2}
\end{center}
\end{figure}

Maeda {\it et al.} \cite{Maeda:13} studied the wavefunctions to extract the correlation between two $\Lambda^*$'s in $K^-K^-pp$. 
As shown in Fig.~\ref{fig:f2}, the tightly bound $K^-p$ units are confined mutually  by a strong binding force, while a similar distribution calculated for $K^-p$ is even more compact than $\Lambda^*\Lambda^*$ in $K^-K^-pp$.

 This indicates that the ensemble of many $K^-$'s and $p$'s tends to form an ensemble of $\Lambda^*$'s. This situation helps diminish possible repulsive interactions among $K^-$'s, as discussed in \cite{Yamazaki:11,Hassanvand:11,Maeda:13}. 
 
One of the important features in kaonic nuclear systems is that the bosonic particle ($K^-$)  produces a molecule-like covalent bonding, called super-strong nuclear force \cite{Yamazaki:07a}. This produces a very large and specially extended force, effectively, by a factor of 4, as strong as the Yukawa-type $N-N$ interaction. This bonding force is taken into account when we calculate multiple $\Lambda^*$ nuclei.  

The experimental data of the binding energy of $K^-pp$ \cite{Yamazaki:10,Ichikawa:15} show about 100 MeV, whereas the theoretical predictions starting from the PDG $\Lambda(1405)$ ansatz, namely, the binding energy of $K^-p$ (= 27 MeV), lead to about 50 MeV binding energy \cite{Yamazaki:02}, which is about half the observed binding energy. As one of the possible reasons we proposed to consider the effect of partial restoration of chiral symmetry in the $\bar{K}N$ interaction, parallel to the case of the $\pi N$ interaction \cite{Suzuki:04,Yamazaki:12}. On the contrary to the "repulsive" $\pi N$ case, the "attractive" $\bar{K} N$ interaction becomes more attractive, causing a very enlarged binding energy, as shown in \cite{Maeda:13}. In the present case, 17 \% enhanced $\bar{K}N$ attraction accounts for the observed binding energy of $K^-pp$. 

\section{$p + p$ reactions to produce $K^-K^-pp$}

We have found from the DISTO experiment at Saturne of Saclay \cite{Yamazaki:10} that $K^-pp$ is abundantly produced from $p + p$ reactions at 2.85 GeV, following a proposed new reaction mechanism  \cite{Yamazaki:07b} of extremely high sticking of $\Lambda^*$ + $p$ that occurs as a doorway state with a short distance of the two participating particles ($\Lambda^*$ and $p$) and contributes to forming only a high-density composite $K^-pp$:
\begin{equation}
p + p \rightarrow p + \Lambda^* + K^+ \rightarrow [p\Lambda^*] + K^+, 
\end{equation} 
\begin{equation}
[p\Lambda^*] \rightarrow K^-pp \rightarrow p + \Lambda.
\end{equation}
Now we can extend this mechanism to produce $K^-K^-pp$ by a direct reaction \cite{Yamazaki:11,Hassanvand:11} as: 
\begin{equation}
p + p \rightarrow  \Lambda^* + K^+ + \Lambda^* + K^+ \rightarrow [\Lambda^* \Lambda^*] + K^+ + K^+, 
\end{equation}
\begin{equation}
[\Lambda^*\Lambda^*] \rightarrow K^-K^-pp \rightarrow  \Lambda + \Lambda.
\end{equation}

\begin{figure}
  \center\includegraphics[width=10cm]{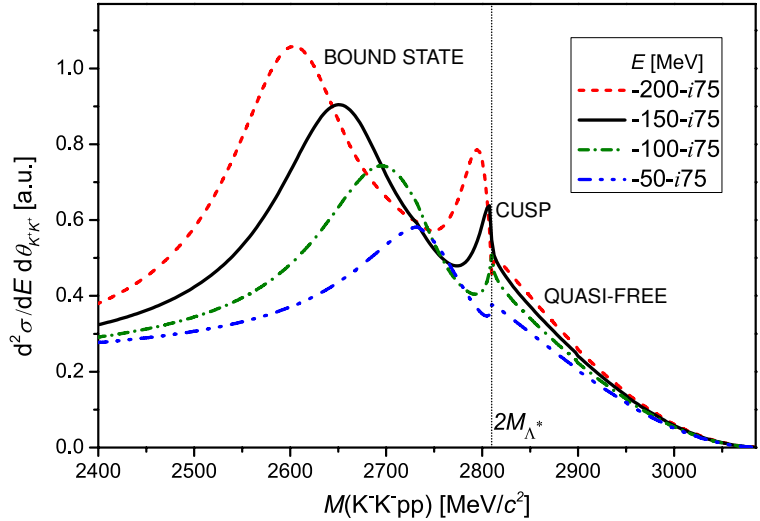}
    \caption{ Differential cross sections for various bound-state energies of the $K^-K^-pp$ system, $E$. For ${T_p}$ = $7.0$ GeV, $b$ = $0.3$ fm and $\theta_{12}$ = $180^{\circ}$. From Hassanvand {\it et al.} \cite{Hassanvand:11}} \label{fig:f4}
    \end{figure}

Detailed calculations for the spectrum shape and cross section were carried out by assuming the real and imaginary parts of the $\bar{K}N$ interaction \cite{Hassanvand:11}. The most suitable incident energy is 7 GeV. The spectrum shapes of $M_{\rm inv}(\Lambda \Lambda)$ are shown in Fig.~\ref{fig:f4}. As in the case of $K^-pp$ production, the intensity of the dominant peak of $M_{\rm inv}(\Lambda \Lambda)$ increases with the growth of the binding energy of $K^-K^-pp$. 

The experimental feasibility of such an experiment should be examined. There is almost no accelerator available in the world that is suitable for this purpose, but there are a few possibilities. In the case of J-PARC, where no primary beam is available at the moment, Sakuma considered to use a secondary proton beam as well as an antiproton beam to evaluate its feasibility  \cite{Sakuma:11}. In the future, if a primary beam of incident energy around 10 GeV can be extracted, the situation will become much better. Of course, there will be many chances for FAIR projects.



\section{Search for $K^-K^-pp \rightarrow \Lambda + \Lambda$ in heavy-ion reactions}

As an extension of the $p+p$ reaction we can conceive Heavy-Ion reactions. which are known to emit substantial amount of $\Lambda$ particles \cite{pbm}. As an extreme case one might challenge to find strange exotica like 
$$K^-K^-pp \rightarrow \Lambda + \Lambda$$
in invariant-mass spectra of $M_{\rm inv}(\Lambda \Lambda)$, which is expected to be around 2.6 GeV/$c^2$.






\begin{thebibliography}{9}
\bibitem{Yamazaki:10} T. Yamazaki {\it et al.}, Phys. Rev. Lett. {\bf 104} (2011) 132502.
\bibitem{Ichikawa:15} Y. Ichikawa {\it et al.},  Prog. Theor. Exp. Phys. {\bf 2015} 021D01.
\bibitem{Akaishi:02} Y. Akaishi and T. Yamazaki, Phys. Rev. C {\bf 65} (2002) 044005. 
\bibitem{Yamazaki:02} T. Yamazaki and Y. Akaishi, Phys. Lett. B {\bf 535} (2002) 70.
\bibitem{Maeda:13} S. Maeda, Y. Akaishi and T. Yamazaki, Proc. Jpn. Acad. Ser. B {\bf 89} (2013) 418.
\bibitem{Barnea:12} N. Barnea, A. Gal and E.Z. Liverts, Phys. Lett. B {\bf 712} (2012) 132.
\bibitem{Yamazaki:11} T. Yamazaki, Y. Akaishi and M. Hassanvand, Proc. Jpn. Acad. Ser. B {\bf 89} (2011) 362. 
\bibitem{Hassanvand:11}M. Hassanvand, Y. Akaishi and T. Yamazaki, Phys. Rev. C {\bf 84} (2011) 015207.
\bibitem{Yamazaki:07a}T. Yamazaki and Y. Akaishi, Proc. Jpn. Acad. Ser. B {\bf 83} (2007) 103. 
\bibitem{Yamazaki:07b}T. Yamazaki and Y. Akaishi, Phys. Rev. C {\bf 76} (2007) 045201. 
\bibitem{Suzuki:04} K. Suzuki {\it et al.}, Phys. Rev. Lett. {\bf 92} (2004) 072302.
\bibitem{Yamazaki:12} T. Yamazaki, S. Hirenzaki, R.S. Hayano, and H. Toki, Phys. Rep. {\bf 514} (2012) 1.
\bibitem{Sakuma:11}F. Sakuma, private communication; also see F. Sakuma {\it et al.}, Hyperfine Interact. (2011)  DOI 10.1007/s10751-011-0393-6.
\bibitem{pbm} P. Braun-Munzinger, this conference.
\end{thebibliography}
\end{document}